\documentclass[a4paper,11pt]{article}
\usepackage{jcappub}
\usepackage[utf8]{inputenc}
\usepackage{amsmath}
\usepackage{amssymb}
\usepackage{graphicx}
\usepackage{xcolor}
\usepackage{multirow}
\usepackage{orcidlink}
\usepackage[normalem]{ulem}

\def\udm{u_{\rm \nu\chi}}

\newcommand{\CLASS}{\texttt{CLASS}}

\title{High resolution Lyman-$\alpha$ forest constraints on dark matter-neutrino scattering}

\author[a,b]{Markus R. Mosbech\orcidlink{0000-0001-7199-6579},}
\author[c,d]{Olga Garcia-Gallego\orcidlink{0009-0009-9003-7889},}
\author[d,e]{Vid Ir\v{s}i\v{c}\orcidlink{0000-0002-5445-461X},}
\author[f,g,h,i]{Matteo Viel\orcidlink{0000-0002-2642-5707},}
\author[b]{Julien Lesgourgues\orcidlink{0000-0001-7627-353X}}

\affiliation[a]{Institute for Theoretical Particle Physics (TTP), Karlsruhe Institute of Technology (KIT), 76128 Karlsruhe, Germany}
\affiliation[b]{Institute for Theoretical Particle Physics and Cosmology (TTK), RWTH Aachen University, D-52056 Aachen, Germany}
\affiliation[c]{Institute of Astronomy, University of Cambridge, Madingley Road, Cambridge CB3 0HA, UK}
\affiliation[d]{KICC - Kavli Institute for Cosmology Cambridge, Madingley Road, CB3 0HA Cambridge, United Kingdom}
\affiliation[e]{Center for Astrophysics Research, Department of Physics, Astronomy and Mathematics, University of Hertfordshire, College Lane, Hatfield AL10 9AB, UK}
\affiliation[f]{SISSA - International School for Advanced Studies, Via Bonomea 265, I-34136 Trieste, Italy}
\affiliation[g]{IFPU, Institute for Fundamental Physics of the Universe, Via Beirut 2, I-34151 Trieste, Italy}
\affiliation[h]{INFN, Sezione di Trieste, Via Valerio 2, I-34127 Trieste, Italy}
\affiliation[i]{INAF - Osservatorio Astronomico di Trieste, Via G. B. Tiepolo 11, I-34143 Trieste, Italy}

\emailAdd{mosbech@physik.rwth-aachen.de}
\emailAdd{og313@cam.ac.uk}
\emailAdd{v.irsic@herts.ac.uk}
\emailAdd{viel@sissa.it}
\emailAdd{lesgourg@physik.rwth-aachen.de}

\abstract{We present new constraints on models of dark matter interacting with neutrino, based on high-resolution Lyman-$\alpha$ forest data. We perform a suite of full hydrodynamical simulations of these models, spanning a range of interaction strengths and thermal histories. We train an emulator on the simulation results. A Monte Carlo Markov Chain analysis yields an upper limit on the interaction strength of $\udm \leq1.5\times10^{-8}$ (95\% C.L.), which is the strongest direct bound to date on such interactions. Our results exclude previous hints of non-zero interactions presented in the literature. We furthermore compare our results to those obtained by mapping warm dark matter constraints to other models with suppressed small-scale structure, and find that these methods would overestimate the constraining power for this model.}

\begin{document}
\begin{flushright}
        {\tt TTP26-29}\\
        {\tt TTK-26-25 }
\end{flushright}	
\maketitle

\section{Introduction}
The $\Lambda$CDM (cosmological constant + cold dark matter) cosmological concordance model has been remarkably powerful in describing observations \cite{Planck:2018vyg}, though with the advent of high precision data, some tensions have arisen between different experiments, in particular regarding the current value of the expansion rate, given by the Hubble parameter $H_0$, and the amplitude of matter clustering, parametrized with $\sigma_8$~\cite{Perivolaropoulos:2021jda}. Recent experiments also begin to show preferences for models beyond the simplest $\Lambda$CDM model, such as the DESI preference for time-evolving dark energy over a cosmological constant~\cite{DESI:2024mwx,Cortes:2024lgw,DESI:2025wyn,Chudaykin:2025lww}.

Dark matter remains an equally mysterious part of our cosmological model. Although it makes up the majority of the matter content of the universe, we do not have a full complete particle physics description of it. Thus far, we know that it clusters on large scales, and have tight upper limits on how strongly it can interact with standard model or light dark sector particles. Dark matter model-building has long been a very active area of research~\cite{Bergstrom:2000pn,Bertone:2004pz,Bertone:2016nfn,Antel:2023hkf,Marsh:2024ury}, while laboratory experiments search for couplings to standard model particles~\cite{CAST:2017uph,Billard:2021uyg,LZ:2024zvo,XENON:2023cxc,ADMX:2023rsk,OHare:2024nmr}. In addition to particle physics experiments, cosmological observations have proven extremely powerful in probing various dark matter properties, including interactions with baryons~\cite{Boehm:2000gq, Chen:2002yh, Boehm:2004th,Dvorkin:2013cea, Dolgov:2013una, CyrRacine:2012fz, Prinz:1998ua,Boddy:2018kfv,Slatyer:2018aqg,Xu:2018efh,Boddy:2018wzy,Becker:2020hzj}, photons~\cite{Boehm:2000gq,  Boehm:2001hm, Boehm:2004th, Sigurdson:2004zp,  McDermott:2010pa,CyrRacine:2012fz, Dolgov:2013una, Wilkinson:2013kia, Boehm:2014vja, Schewtschenko:2014fca, Schewtschenko:2015rno, Ali-Haimoud:2015pwa,  Escudero:2015yka,  Diacoumis:2017hff, Stadler:2018jin,  Stadler:2018dsa,Lopez-Honorez:2018ipk,Becker:2020hzj}, neutrinos~\cite{Boehm:2000gq, Boehm:2004th, Mangano:2006mp, Serra:2009uu, Wilkinson:2014ksa, Ali-Haimoud:2015pwa, DiValentino:2017oaw, Stadler:2019dii, Mosbech:2020ahp,Hooper:2021rjc,Akita:2023yga,Brax:2023rrf,Giare:2023qqn,Mosbech:2024wxr,Zu:2025lrk,Zhou:2026lqt}, dark radiation species~\cite{Kaplan:2009de, Das:2010ts, Diamanti:2012tg,  Buen-Abad:2015ova, Lesgourgues:2015wza, Das:2017nub, Ko:2017uyb, Escudero:2018thh,Archidiacono:2019wdp,Becker:2020hzj,Plombat:2024kla}, DM self-interactions~\cite{Carlson:1992fn, deLaix:1995vi, Spergel:1999mh, Dave:2000ar, Creasey:2016jaq, Rocha:2012jg, Kim:2016ujt, Huo:2017vef, Markevitch:2003at, Randall:2007ph, Boehm:2000gq, Boehm:2004th}. Cosmological data also constrain models of warm~\cite{Dodelson:1993je,Bode:2000gq,Hansen:2001zv,Asaka:2005an,Viel:2005qj,Boyarsky:2009ix,Viel:2013fqw,Abazajian:2001nj,Boyarsky:2008xj,Dolgov:2000ew,Irsic:2023equ,Garcia-Gallego:2025kiw} and axionic~\cite{Hui:2021tkt,Armengaud:2017nkf,Irsic:2017yje,Lidz:2018fqo,Rogers:2020ltq} DM.

Some recent studies suggest that current CMB and structure formation data may favour a non-zero coupling between dark matter and neutrinos \cite{Hooper:2021rjc,Giare:2023qqn,Zu:2025lrk}. Thus, this scenario is of particular interest. Dedicated hydrodynamical simulations are necessary to extend the observational tests of this model to non-linear scales. In this work, we perform a suite of such simulations. We then compare the model predictions on mildly non-linear scales to measurements of the Lyman-$\alpha$ flux power spectrum from high-resolution quasar spectra.

The paper is structured as follows: In Section~\ref{sec:model} we present the interacting dark matter model itself; In Section~\ref{sec:simulations} we describe our simulations; In Section~\ref{sec:data} we describe the Lyman-$\alpha$ forest data used in our analysis; In Section~\ref{sec:results} we present our results; In Section~\ref{sec:discussion} we discuss how they fit in the literature and compare them with bounds deribed from other methods; We present our conclusions in Section~\ref{sec:conclusion}.

\section{Dark matter-neutrino interactions}
\label{sec:model}

The scattering of heavy dark matter particles with hot, light particles such as neutrinos suppresses clustering on scales smaller than the ``collisional damping scale'' set by the interaction~\cite{Boehm:2004th}. In general, dark matter and neutrinos are coupled in the early universe, similarly to the baryon-photon fluid prior to recombination. The pressure from hot neutrinos prevents gravitational collapse until the two species decouple. The sound horizon at this ``dark decoupling'' sets the collisional damping scale. The pressure of the coupled fluid also induces ``dark acoustic oscillations'' below this scale, similar to photon-baryon acoustic oscillations in the visible sector~\cite{Cyr-Racine:2013fsa,Cyr-Racine:2015ihg}. Other models of interaction between dark matter and relativistic species, such as photons or dark radiation, have a qualitatively similar impact on dark matter clustering~\cite{Murgia:2017lwo,Boehm:2000gq,Ali-Haimoud:2015pwa,Stadler:2018jin,Becker:2020hzj,Ackerman:2008kmp,Cyr-Racine:2015ihg,Archidiacono:2019wdp,Hooper22,Plombat:2024kla}. Models with interactions between dark matter and neutrinos are particularly interesting to investigate, both from the particle physics perspective, as the properties of neutrinos are not yet fully understood, and from the observational side, as recent analyses have indicated a preference for a non-zero interaction strength in observational data~\cite{Hooper:2021rjc,Giare:2023qqn,Zu:2025lrk}. Furthermore, these models are very difficult to probe in laboratory experiments. Cosmology offers a unique opportunity to test them.

We parametrize the interaction in terms of the effective interaction strength parameter introduced in Ref.~\cite{Boehm:2001hm},
\begin{equation}    u_{\nu\chi}\equiv\frac{\sigma_{\nu \chi}}{\sigma_\mathrm{Th}}\left(\frac{m_\chi}{100\,\mathrm{GeV}}\right)^{-1},
\end{equation}
where $\sigma_\mathrm{Th}\approx6.65\times10^{-29}\,\mathrm{m}^2$ is the Thomson scattering cross-section, $m_\chi$ is the dark matter particle mass, and $\sigma_{\nu\chi}$ is the elastic scattering cross section between interacting DM particles and neutrinos. We use this parameterisation because of the degeneracy between the scattering cross section and dark matter mass in the scattering rate. 
Like in most of the literature on DM-neutrino interactions, including the recent studies suggesting a non-zero coupling, we assume that $\sigma_{\nu\chi}$ is constant in time, although realistic models could also feature a cross-section with a smooth temperature dependence \cite{Olivares-DelCampo:2017feq} or enhanced resonantly over a narrow redshift range \cite{Trojanowski:2025oro}.   
The rate expressed with respect to conformal time reads
\begin{equation}
    \dot{\mu}\equiv a \sigma_{\nu \chi} n_\chi = \frac{\sigma_\mathrm{Th}}{100 \, \mathrm{GeV}} a u_{\nu\chi} \rho_\chi,
\end{equation}
with $n_\chi$ being the interacting DM particle number density, and $\rho_\chi$ its energy density. In the presence of scattering and in the Newtonian (aka longitudinal) gauge, the linear Euler equations for neutrinos and dark matter are given as~\cite{Boehm:2001hm,Mangano:2006mp,Serra:2009uu,Wilkinson:2014ksa,DiValentino:2017oaw,Stadler:2019dii}
\begin{subequations}
\begin{align}
    \dot{\theta}_\nu &= k^2\psi + k^2 \left(\frac{1}{4}\delta_\nu - \sigma_\nu\right) - \dot{\mu}\left(\theta_\nu - \theta_\chi\right),\\
    \dot{\theta}_\chi &= k^2 \psi - \mathcal{H}\theta_\chi - \frac{4\rho_\nu}{3\rho_\chi}\dot{\mu}\left(\theta_\chi - \theta_\nu\right).
\end{align}
\end{subequations}

We use a version of \CLASS{} modified to include these interactions to compute the linear matter power spectrum, from which we generate the initial conditions for our hydrodynamical simulations. In our simulations, we choose the interaction strengths such that the exponential suppression in the power spectrum arises approximately at the same scale as for thermal warm dark matter with masses 1.5, 2, 3, and 4 keV respectively, since this is the regions of parameter space where we expect to find constraints from the Lyman-$\alpha$ forest.

\begin{figure}
    \centering
    \includegraphics[width=0.6\linewidth]{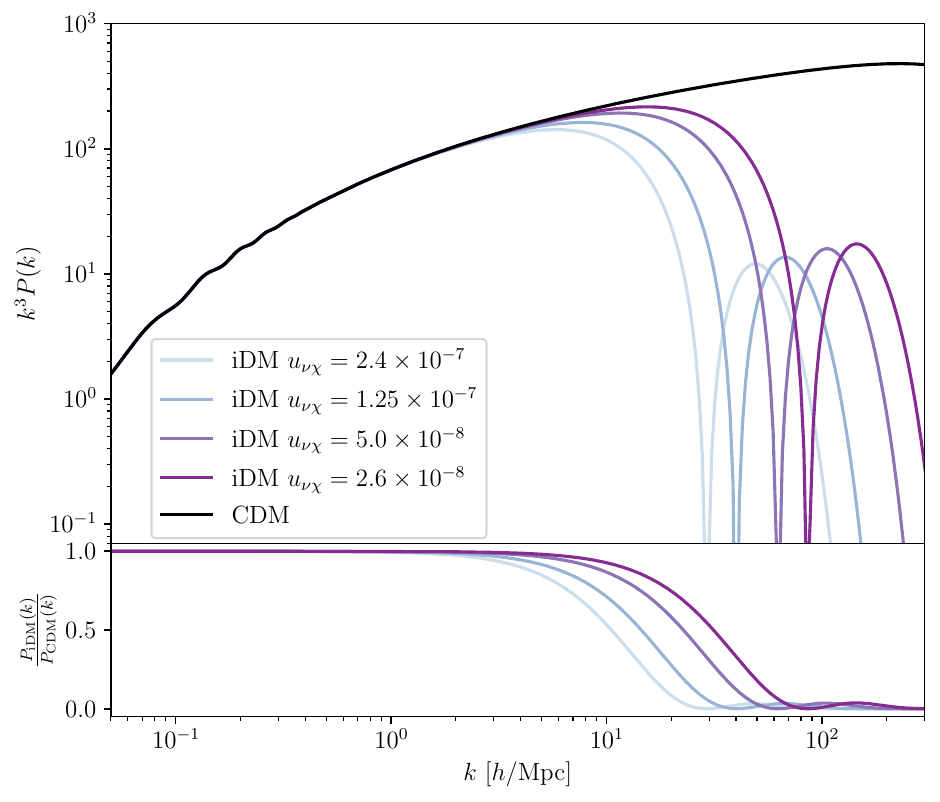}
    \caption{Linear power spectra at redshift $z=0$ for the DM-neutrino interaction models considered in our simulations, and their ratios with respect to the power spectrum of an equivalent model with decoupled CDM.}
    \label{fig:linear}
\end{figure}

For the sake of simpler and faster hydrodynamical simulations, we approximate neutrinos as massless particles. This does not qualitatively affect the results, as the inclusion of masses has been shown to have only a minor effect on the final power spectrum~\cite{Mosbech:2020ahp,Giare:2023qqn}.

\section{Simulations}
\label{sec:simulations}
We perform a suite of dedicated simulations using the \texttt{P-Gadget3} code based on the Sherwood-Relics suite \cite{Puchwein:2023}. The initial conditions are generated with \texttt{N-GenIC} from linear matter power spectra computed with \CLASS{} at a redshift of $z=99$. The simulations use $2\times 1024^3$ dark matter and gas particles and a box volume of $\left(20\, \text{Mpc}/h\right)^3$. To guarantee numerical convergence, we also perform a few supplemental simulations with varying mass resolution, $R_{s}$, and box size, as discussed in Ref.~\cite{Irsic:2023equ}. 

Our grid of simulated models includes modifications to the spatially uniform UV background described in Ref.~\cite{Puchwein:2019}, which specifies the reionization history. The timing and duration of reionization affect the hydrodynamic response of the gas to heating \cite{Gnedin:1998}. This effect can be described through the cumulative energy per proton deposited into the gas by photo-heating at redshift $z$, $u_0(z)$ \cite{Nasir:2016}. We vary the end of reionization redshift, $z_{\text{rei}}^{{\text{end}}}$,  to simulate different integrated thermal histories, and therefore a different evolution of $u_0(z)$. For each reionization model, we further rescale the photo-heating rates in the same way as in Ref.~\cite{Puchwein:2019}, which gives three models for each $z_{\text{rei}}^{{\text{end}}}$, with three different gas temperatures. Therefore, for each of our four choices of iDM interaction rate $\udm$, we run 12 different thermal histories, identical to those presented in Table 1 of Ref.~\cite{Irsic:2023equ}. This yields a total of 48 simulations. 

To avoid wasting computing time, we reuse the CDM simulations of Ref.~\cite{Garcia-Gallego:2025kiw} for the CDM reference case, that is, the limiting case with $\udm = 0$. We only perform one CDM validation simulation to cross-check that our set-up has not changed with respect to Ref.~\cite{Garcia-Gallego:2025kiw}.

From the simulation output, the optical depth $\tau(z)$ along the line-of-sight (LOS) skewers allows to construct our estimator, the 1D flux power spectrum in each redshift bin, $P_{\text{F}}(k,z)$ = $L^{-1}_{\text{box}}{\langle{\delta_{\text{F}}}{\delta^{*}_{\text{F}}}\rangle}$, with flux fluctuations $\delta_{\text{F}}=F/{\bar F}-1$, with $\bar{F}$ corresponding to the averaged mean transmitted flux across the LOS pixels. As in Ref.~\cite{Irsic:2023equ}, we estimate $P_\mathrm{F}$ by extracting the flux from 5000 different lines of sight through the simulation box. Each flux is Fourier transformed before averaging over the lines of sight.
Similarly to Ref.~\cite{Irsic:2023equ}, we largely increase the number of models in the post-processing phase by translating and rotating the particle distribution in the temperature-density plane along the LOS. This effectively isolates the impact of thermal broadening under the assumption of photo-ionization equilibrium, which allows us to parametrize the temperature-density relation in terms of an instantaneous gas temperature at mean density, $T_0$, and a power-law index $\gamma$, see Ref.~\cite{Boera:2019}. During the post-processing, we further adjust the mean transmission $\bar{F}$ to match the IGM transmission opacity $\tau_{\rm{eff}}$ from \cite{Boera:2019} through a rescaling of the $\tau$ field. The range spanned by the free parameters ($T_0$, $\gamma$, $\tau_{\rm{eff}}$) at each redshift bin is discussed in Ref.~\cite{Garcia-Gallego:2025kiw}.

\section{Data}
\label{sec:data}

The simulations are compared with the flux power spectra presented in Ref.~\cite{Boera:2019}, which are estimated from HIRES/Keck and UVES/VLT high-resolution observations of 15 quasars. The data set consists of measurements of the 1D flux power spectrum at $z_{\rm{bin}}$ = 4.2, 4.6, and 5.0. It extends to scales twice as small as those resolved by previous Lyman-$\alpha$ flux power spectrum analyses (e.g. \cite{Irsic:2017}), reaching $k_{\rm{max}}$ = 0.2 km$^{-1}$s. This improvement is mainly due to a more rigorous modeling of the systematics affecting the smallest scales (see \cite{Irsic:2023equ} for more details). This data set yields the tightest constraints to date on a range of dark matter models, including WDM \cite{Irsic:2023equ}, which motivates its use in our analysis. 
We compare the measured and simulated power spectra, $P_{\text{F}}(k,z)$, using the same set-up as in \cite{Garcia-Gallego:2025kiw}, namely, the Monte Carlo Markov Chain (MCMC) Bayesian framework implemented in \texttt{Cobaya} \cite{Torrado:2020dgo,Cobaya}. 

\begin{figure}[hbtp!]
    \centering
    \includegraphics[width=0.7\linewidth]{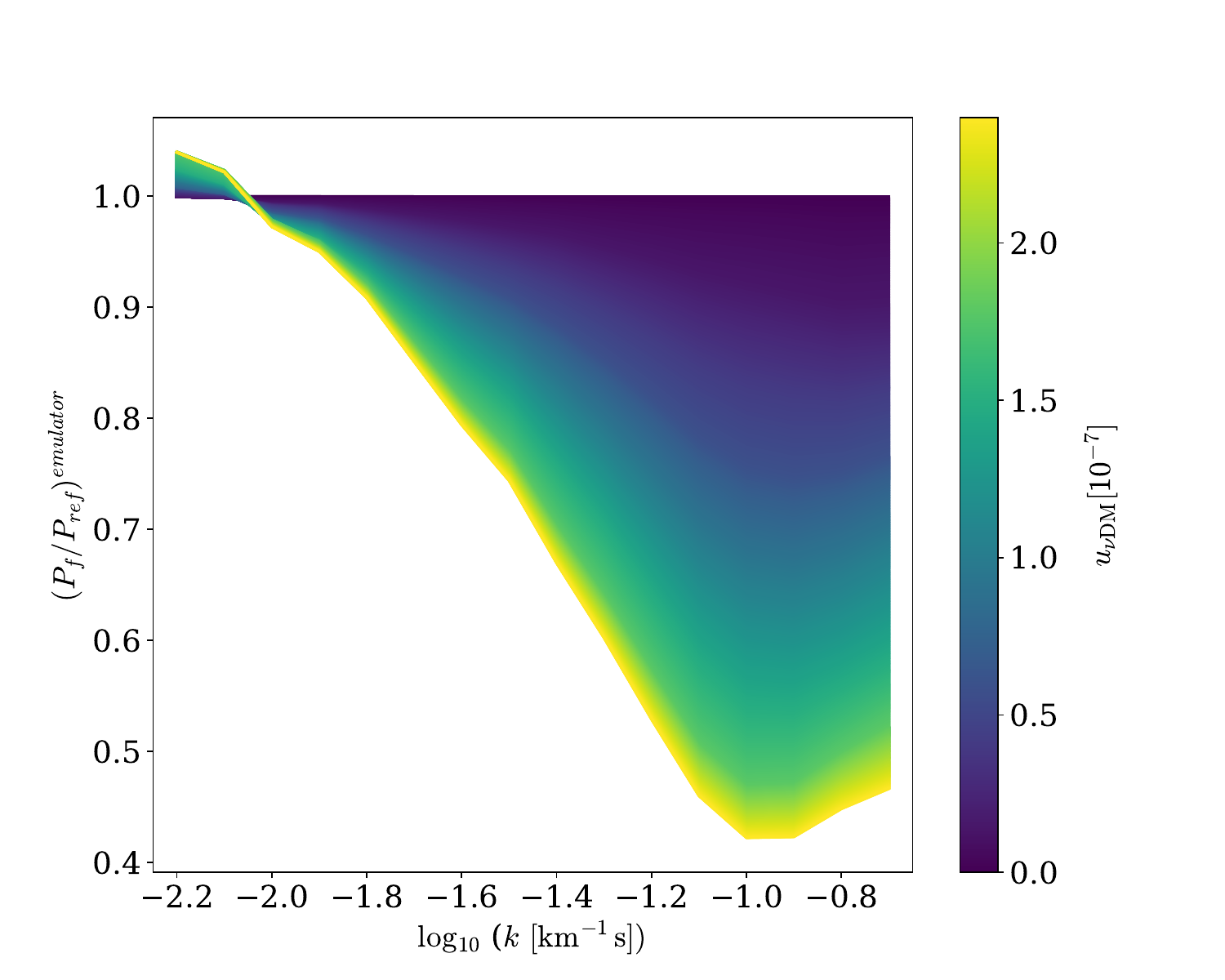}
    \caption{Ratio of predicted models within $\udm$ grid to reference (CDM) model at $z$=4.2, fixing the other parameters that affect $P_{\rm{F}}$. }
    \label{fig:pred}
\end{figure}

To get fast theoretical predictions, we train a neural network (NN) emulator of the flux power spectrum on our suite of simulations, to cover the full range of scanned thermal and DM parameters. Our new emulator shares the same characteristics as the one presented in \cite{Garcia-Gallego:2025kiw} for the case of pure WDM models. This is justified by the fact that the two DM models have a similar imprint on the flux spectrum, while the two simulation suites have a similar size and structure. The emulator is then used to evaluate the likelihood of the data given the model following Ref.~\cite{Boera:2019}.
The main characteristics of the emulator include a three-layer NN architecture of $[60, 60, 60]$ neurons, an initial learning rate hyperparameter \texttt{lr=$10^{-3}$}, adaptive updates based on the validation loss using the \texttt{Adam} optimization algorithm, and a batch size of 12.  We verify that the emulator prediction on the hidden test set, representing 10\% of the complete dataset, differs from the true input by $<1.5\%$. We further verify that the emulator provides smooth predictions across the parameter space. In Figure ~\ref{fig:pred}, we show the ratio of the emulator predictions to the corresponding CDM models for thousand models at $z$=4.2, varying $\udm$ while keeping the remaining parameters ($T_0, \gamma, \tau_{\rm{eff}}, u_0$) fixed.

The robust recovery of synthetic spectra $P_{\text{F}}(k,z)$ using the emulator allows us to significantly speed up the evaluation of the likelihood, and thus, the inference of parameters.

\section{Results}
\label{sec:results}

\begin{figure}
    \centering
    \includegraphics[width=0.6\linewidth]{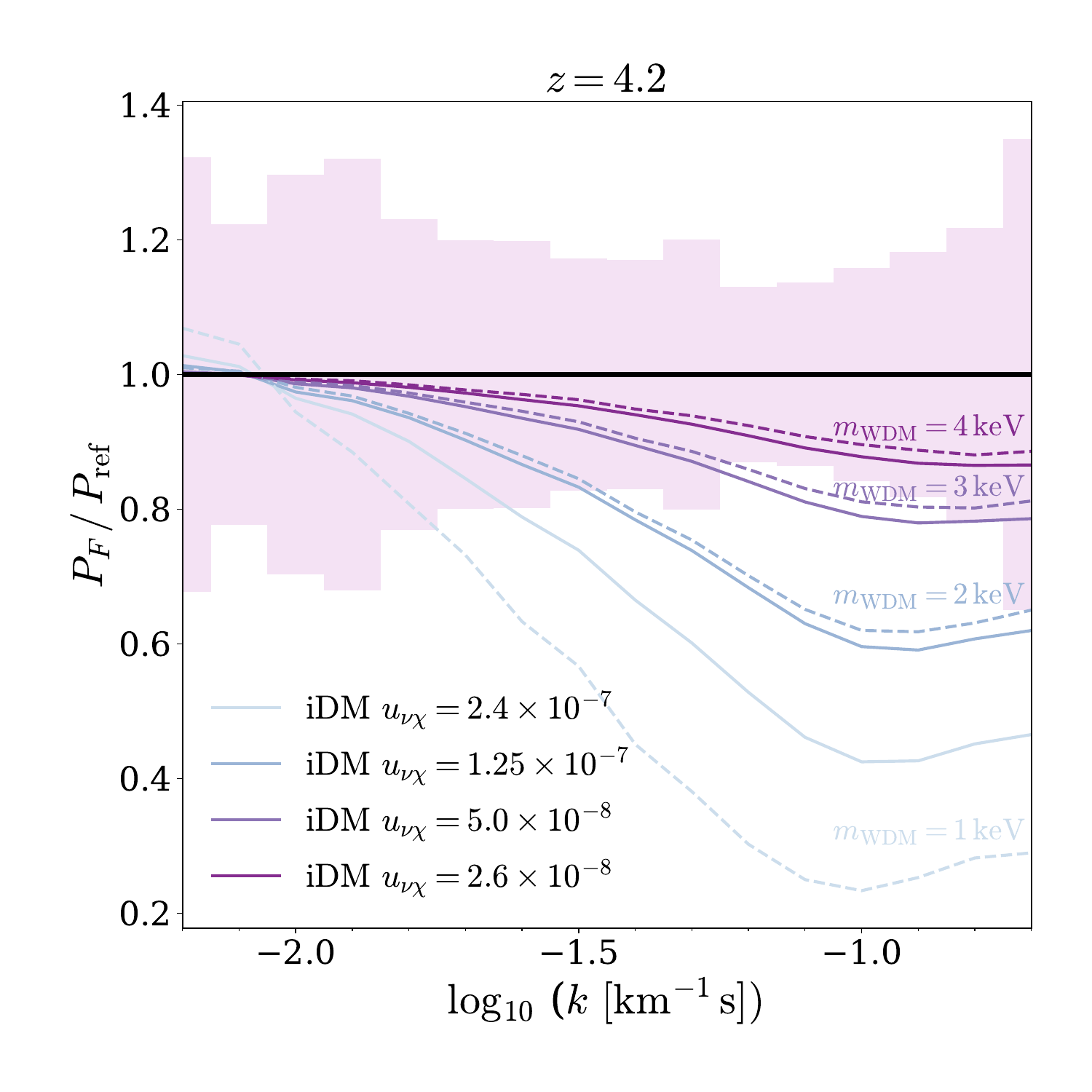}
    \caption{The solid lines show the flux power spectrum extracted from simulations for various interaction rates $\udm$ divided by that from an equivalent CDM model ($\udm=0$). For comparison, the dashed lines show the same ratios for the WDM models used in the simulation suite of Ref.~\cite{Garcia-Gallego:2025kiw}, with initial conditions generated from smooth power spectra without oscillations.}
    \label{fig:fluxpk_z4.2}
\end{figure}

The predicted Lyman-$\alpha$ flux power spectrum that we extract from our simulations with DM-neutrino interactions shows a suppression of power on small scales similar to that observed in warm dark matter models. In particular, we find that the strong oscillations present in the input spectra used for generating initial conditions (see Fig.~\ref{fig:linear}) are completely washed out in the flux spectrum (see Fig.~\ref{fig:fluxpk_z4.2}). This matches expectations from gravity-only N-body simulations, which indicate that non-linear structure formation washes out such features in the non-linear matter power spectrum of this family of models by redshifts $z\lesssim10$~\cite{Mosbech:2022uud}. Our simulations confirm that this also holds true for the Lyman-$\alpha$ flux spectrum, which is a probe of regions of intermediate density rather than strongly non-linear high-density environments \cite{Weinberg:1998}. The features are further washed out in the 1D Lyman-$\alpha$ forest flux power spectrum due to the projection effect in the mapping from 3D to 1D. The oscillations in this family of models tend to create an overall amplitude suppression effect, rather than preserve the oscillatory pattern \citep{Bose:2018juc}.

Our MCMC analysis constrains the 
astrophysical parameters ($T_0$, $\gamma$, $u_0$, $\tau_{\rm{eff}}$) in each redshift bin, as well as the redshift-independent cosmological parameter $\udm$. We denote the redshift-dependent parameters with a superscript, e.g., $T_0^{5.0}\equiv T_0(z_\mathrm{bin}=5.0)$. For the astrophysical parameters, we adopt the same priors as in Ref.~\cite{Garcia-Gallego:2025kiw}. For the interaction rate, we assume a uniform prior in the range $\udm \in [0, 2.4] \times 10^{-7}$. This range matches the values adopted in our simulation suite and in the emulator training. Thus, we avoid any extrapolation beyond the simulated range.

The astrophysical parameters that can suppress $P_{\rm{F}}(k,z)$ on small scales and are potentially degenerate with $\udm$ in each redshift bin are $u_0$ and $T_0$. Thus, the choice of prior for $(u_0, T_0)$ has a significant impact on our final bound on $\udm$. In principle, since the different thermal histories assumed in each simulation were chosen on the basis of observational constraints, the band spanned by these models already imposes a prior in the $u_0-T_0$ plane in each bin, as explored by Ref.~\cite{Irsic:2023equ}. However, we choose to investigate the impact of two (stronger and more informative) priors based on independent observations:
\begin{itemize}
\item \textbf{Gaussian} $\boldsymbol\tau_\mathbf{CMB}$ \textbf{prior.}
First, we impose a thermal prior based on the relation between the reionization optical depth and $(u_0^{5.0}, T_0^{5.0})$, inferred in  Ref.~\cite{GarciaGallego:2025b} from the coupling between the thermal state and the ionization state of the IGM, described by the ionization fraction $x_e(z)$ or optical depth $\tau(z)$. We use the CMB constraint on the reionization optical depth, $\tau_{\rm{CMB}}= 0.054\pm0.007$ \cite{Planck:2018vyg}, as an implicit prior on $(u_0^{5.0}, T_0^{5.0})$. Since the range of $(u_0, T_0)$ values adopted in our simulations is designed to provide realistic reionization histories that are roughly compatible with CMB bounds, the $\tau_\mathrm{CMB}$ prior selects a region in the $(u_0^{5.0}, T_0^{5.0})$ plane which is qualitatively similar to the band spanned by our simulations and covered by our emulator. However, imposing a Gaussian prior on $\tau_\mathrm{CMB}$ rather than a uniform prior within this band is more constraining. In other redshift bins, we stick to a uniform prior in the range of values of $(u_0^{4.2}, T_0^{4.2})$ and $(u_0^{4.6}, T_0^{4.6})$ spanned by our simulations.
\item \textbf{Gaussian} $\mathbf{T_0}$ \textbf{prior.} As an alternative, we perform an analysis with Gaussian priors on $T_0 ^{z_{\rm{bin}}}$ in each bin, based on the $T_0(z)$ model presented in Ref.~\cite{Irsic:2023equ}, which rescales the fiducial thermal history from Ref.~\cite{Puchwein:2019} to fit independent $T_0$ measurements from Refs.~\cite{Gaikwad:2020, Gaikwad:2021} at higher and lower redshifts. The model from Ref.~\cite{Irsic:2023equ} therefore provides  independent $T_0 ^{z_{\rm{bin}}}$ priors at the central values of the Lyman-$\alpha$ data from Ref.~\cite{Boera:2019}. We then stick to uniform priors on the $u_0 ^{z_{\rm{bin}}}$ parameter. 
\end{itemize}
The remaining parameters $\gamma^{z_{\rm{bin}}}, \tau_{\rm{eff}}^{z_{\rm{bin}}}$ are sampled with uniform priors over the range spanned by  the simulated models. Since they affect $P_{\rm{F}}(k,z)$ in a scale-independent way, they have no impact on the $\udm$ constraints. We cross-checked that in the decoupled CDM limit and with each choice of priors, we are able to reproduce the results of Ref.~\cite{Garcia-Gallego:2025kiw}.

\begin{figure} 
    \centering
    \includegraphics[width=0.5\linewidth]{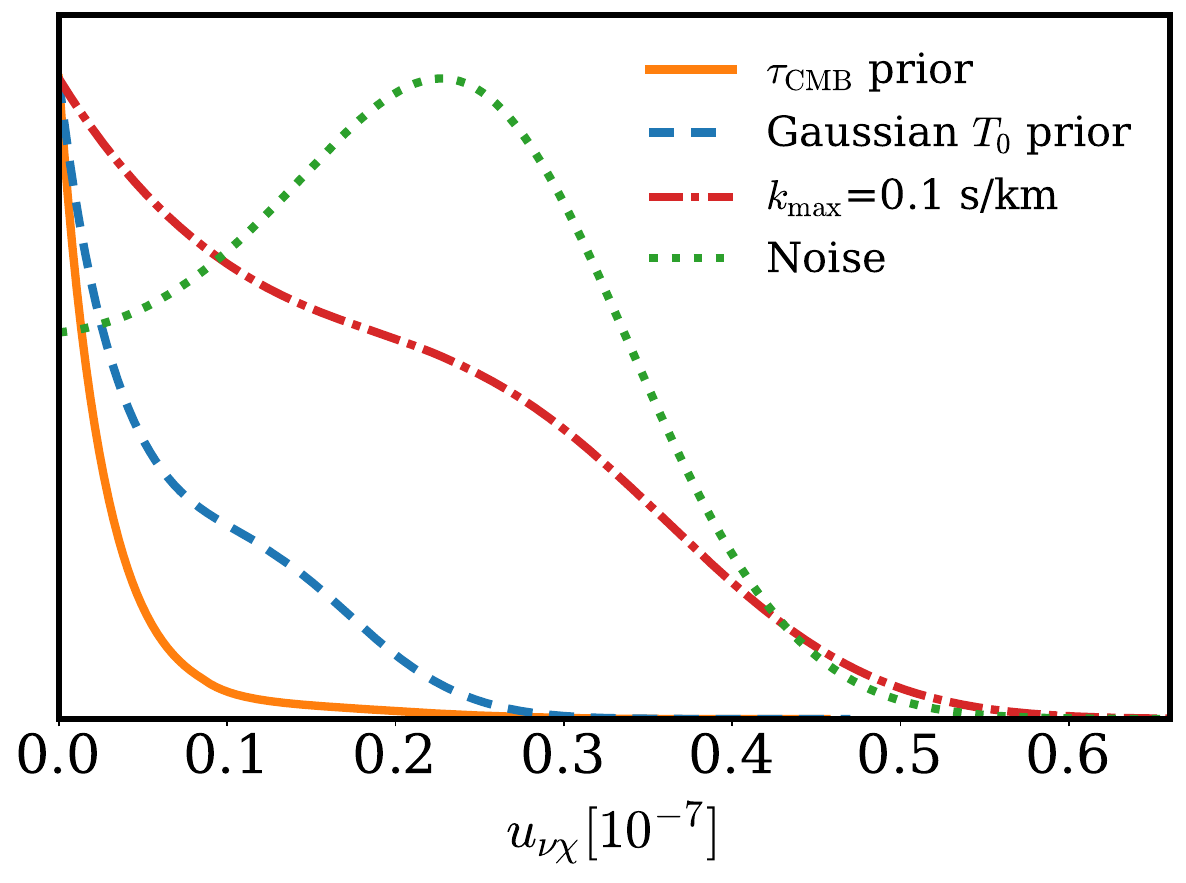}
    \caption{1D posterior distribution for $\udm$ for the different MCMC analyses with different thermal priors (solid and dashed), without the last three $k$-bins (dash-dotted), and with an increased instrumental noise level (dotted).}
    \label{fig:1d}
\end{figure}

The $\tau_{\rm{CMB}}$ prior provides our baseline result, shown as the solid orange posterior in Fig.~\ref{fig:1d} (with label `$\tau_{\rm{CMB}}$ prior'). This yields our most stringent upper limit on $\udm < 1.5 \times 10^{-8}$ (95\% C.L). In this case, we stick to loose priors on the parameters $(u_0^{4.2}, u_0^{4.6})$, which are left largely unconstrained by the data (consistently with Figure 7 in \cite{Irsic:2023equ}). Due to parameter degeneracies and an anti-correlation between $u_0^{z_\mathrm{bin}}$ and $T_0^{z_\mathrm{bin}}$, this pushes $(T_0^{4.2}, T_0^{4.6})$ to stay close to the lower limit of their prior (see Fig.~\ref{fig:corner} in Appendix~\ref{sec:triangle}). Consequently, this leads to tight constraints on $\udm$.

The $T_0$ prior results in a looser bound on the DM-neutrino interaction strength, $\udm<2.03 \times 10^{-8}$ (95\% C.L), corresponding to the dashed-blue posterior (with label `Gaussian $T_0$ prior') in Fig.~\ref{fig:1d}. Compared to the results inferred from the $\tau_{\rm{CMB}}$ prior, this case favours larger values of $T_0^{z_\mathrm{bin}}$ and thus smaller values of $u_0^{z_\mathrm{bin}}$ (see Fig.~\ref{fig:corner} in Appendix~\ref{sec:triangle}). This leaves more room for a possible suppression of the small-scale flux power spectrum with $\udm$. The goodness of fit is better in the case of the $\tau_{\rm{CMB}}$ prior ($\chi^2_{\rm min}=34.3/35$) than with the $T_0$ prior ($\chi^2_{\rm min} =47/35$).

Using $\tau_{\rm{CMB}}$ as our thermal prior baseline, we run two additional analyses to assess the robustness of our results. We first focus on the impact of the last three $k$-bins on the constraining power for $\udm$. Following Ref.~\cite{Irsic:2023equ}, we remove the data points with $k > 0.1 \, {\rm s}/{\rm km}$, that is, extending beyond the range of previous data sets used in Lyman-$\alpha$ forest analyses \cite{Viel:2013fqw, Irsic:2017}. We find that the constraint weakens to $\udm < 4.1 \times 10^{-8}$ (95\% C.L, dash-dotted line labelled `$k_\mathrm{max} \leq 0.1 \, {\rm s}/{\rm km}$' in Fig.~\ref{fig:1d}). A similar softening of the constraints was found in the pure WDM case \cite{Garcia-Gallego:2025kiw}. However, we highlight that even in the absence of the smallest-scale data points, our constraints remain two orders of magnitude more stringent than previous limits, e.g., from Ref.~\cite{Hooper:2021rjc}.

We further run an extremely conservative analysis in which we artificially increase the instrumental noise level beyond the already mean-subtracted noise described in Ref.~\cite{Boera:2019}. Our goal is to check how much the $\udm$ bound degrades if the instrumental noise turns out to be underestimated. To achieve this, we add to the flux power spectrum returned by the emulator a term that quantifies the deviation of the noise flux power from its mean, following the noise distribution modelled by \cite{Irsic:2023equ}. This results in $\udm < 3.9\times 10^{-8}$ (95\% C.L), as shown by the dotted line labelled `Noise' in Fig.~\ref{fig:1d}. In principle, a few other effects could lead to underestimated systematics, such as patchy reionization corrections and thermal-dependent resolution corrections. However, Ref.~\cite{Garcia-Gallego:2025kiw} has shown that these effects are subdominant and only likely to change the parameter constraints by 5-10\%.

\begin{figure} 
    \centering
    \includegraphics[width=\linewidth]{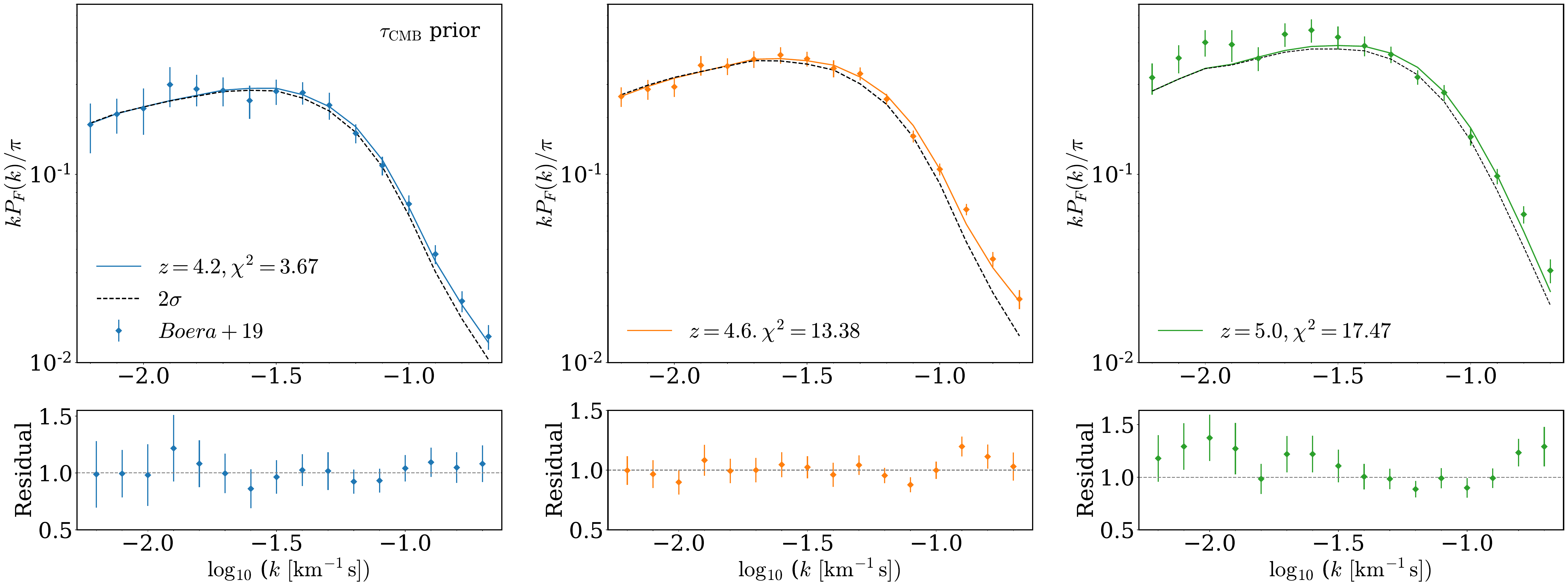}
    \caption{Best-fit flux power spectrum (solid lines) and residuals (lower panels) at each redshift bin for the analysis with $\tau_{\rm{CMB}}$ priors. This model has a vanishingly small interaction rate, $\udm=1.8\times 10^{-11}$. The contribution of each bin to the minimum $\chi^2$ are also indicated (they add up to $34.4/35$). We further show the flux power spectrum for the same parameters but $\udm = 1.5 \times 10^{-8}$ (dashed lines), which corresponds to our upper 95\% C.L.}
    \label{fig:bestfit}
\end{figure}

Fig.~\ref{fig:bestfit} shows the flux spectrum of our best-fit model (with a vanishingly small value of $\udm$) compared to that of a model with the same parameters but $\udm=1.5\times10^{-8}$ (which corresponds to our 95\% C.L. upper bound with the $\tau_{\rm{CMB}}$ prior). It is clear from the figure that the non-interacting model is a better fit to the data.

\section{Discussion}
\label{sec:discussion}

Our results present an interesting contrast to recent analyses of the DM-neutrino interactions based on Lyman-$\alpha$~\cite{Hooper:2021rjc}, CMB~\cite{Giare:2023qqn} and large scale structure~\cite{Zu:2025lrk} data, which have shown a mild preference for a non-zero interaction strength $\udm$. These analyses favour values close to $\udm \sim5\times 10^{-6}$ for Lyman-$\alpha$~\cite{Hooper:2021rjc} (massive neutrinos), in the range $[2,4]\times10^{-5}$ (massless neutrinos) or $[ 2\times 10^{-6}, 10^{-4} ]$ (massive neutrinos) for different combinations of CMB+BAO experiments~\cite{Giare:2023qqn}, and close to $2\times10^{-4}$ (massless neutrinos) when combining CMB+BAO with DES Y3 cosmic shear data~\cite{Zu:2025lrk}. These are all robustly excluded by several orders of magnitude in our analysis, even under our weakest assumptions for the astrophysical parameters or when removing the smallest-scale bins, as shown in Fig.~\ref{fig:1d}.

Our constraints with both thermal priors are the strongest to date inferred from direct simulations of the interacting DM-neutrino model. They improve upon limits from the galaxy luminosity function from Ref.~\cite{Mosbech:2022nkk}, which found that it was impossible to reproduce the observed luminonity function for values of $\udm$ of $10^{-6}$ and larger. A recent work on Milky Way satellites~\cite{Crumrine:2024sdn} puts a limit comparable to that obtained in our analysis. However, the authors of Ref.~\cite{Crumrine:2024sdn} performed simulations for WDM and matched the suppression scale of WDM and iDM models, rather than directly simulating the iDM scenario. 
Using observations by IceCube of high-energy neutrinos coming from the active galactic nuclei (AGN) of NGC 1068, Ref.~\cite{Cline:2023tkp} was also able to put a bound on the cross-section comparable to ours, but subject to different astrophysical uncertainties on the modeling of the dark matter halo and of active nucleus of NGC 1068. Other constraints derived from the observation of other objects by IceCube are less stringent \cite{Zapata:2025huq}.
We also notice that, according to the sensitivity forecast of the 21cm-line intensity-mapping from SKA reported in Refs.~\cite{Mosbech:2022uud,Dey:2022ini}, our bounds are also stronger than what can be achieved with SKA.

\begin{figure} 
    \centering
    \includegraphics[width=\linewidth]{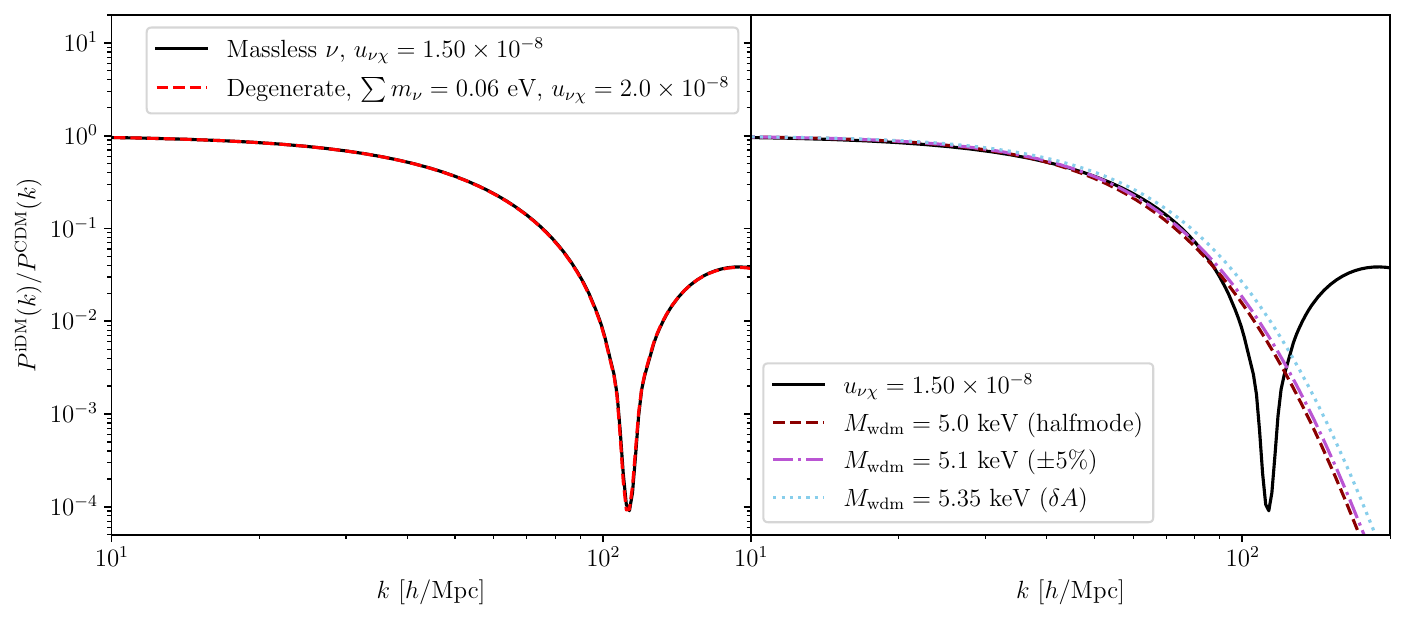}
    \caption{\emph{Left:} Comparison of the linear matter power spectrum suppression for a model of DM interacting with massless neutrinos and another model of DM interacting with three neutrinos degenerate in mass with $\sum m_{\nu}=0.06$ eV. The interaction rate of the second model has been adjusted to obtain the same small-scale suppression. \emph{Right:} Comparison of the linear matter power spectrum suppression for an iDM model with $\udm=1.5\times 10^{-8}$, corresponding to our upper 95\% C.L., and three WDM models in which the equivalent WDM mass is identified via the $\pm 5\%$, half-mode, or $\delta A$ criteria.}
    \label{fig:mnu+wdm}
\end{figure}

Our initial conditions were produced under the massless neutrino approximation for consistency with the set up of our hydrodynamical simulations. Since the interaction strength parameter $\udm$ only enters via the initial conditions, one can match the linear matter power spectrum in the massless case to one in the massive case with a similar suppression scale to obtain an estimate of the bound in the latter case. For instance, if we assume three neutrinos degenerate in mass with $\sum m_{\nu}=0.06$ eV, this approximate method gives $\udm \leq2.0\times10^{-8}$. As shown in the left panel of Fig.~\ref{fig:mnu+wdm}, the matching is nearly perfect at the level of the linear matter power spectrum. We thus expect the estimated bound to be robust.

We also perform a comparison with other approximate methods proposed in the literature to assign an equivalent WDM mass to any non-standard DM model with suppressed small-scale structure, for the purposes of constraining such models without having to perform dedicated simulations. Our 95\% C.L. limit $\udm = 1.5\times 10^{-8}$ maps to three different WDM masses when using three different criteria suggested in earlier works. Firstly, the $\pm 5\%$-criterion (maximising the $k$-scale to which the iDM model linear $P(k)$ matches that of thermal WDM to within $\pm5\%$~\cite{Mosbech:2022uud}), which we used in this work to identify relevant values of $\udm$ to use in our simulation grid, yields an equivalent WDM mass $m_\mathrm{wdm}=5.1\,{\rm keV}$. Secondly, the half-mode criterion as described in e.g. Refs.~\cite{Crumrine:2024sdn,Nadler:2025fcv}, based on matching the half-mode scale $k_{1/2}$ where $P^{\rm extended}/P^{\Lambda \text{CDM}}=0.25$, gives $m_\mathrm{wdm}=5.15\,{\rm keV}$. Finally, the $\delta A$ criterion of Ref.~\cite{Murgia:2017lwo} implies $m_\mathrm{wdm}=5.35\,{\rm keV}$.
The linear power spectra of the three equivalent WDM models are compared to that of the original iDM model in the right panel of Fig.~\ref{fig:mnu+wdm}.
The three equivalent WDM masses are relatively close to each other, but all three methods would lead to an overestimation of the constraining power of Lyman-$\alpha$ data on exotic dark matter models, since a dedicated WDM analysis based on the same data set and pipeline as this work yields $m_\text{wdm} > 5.86$ keV (95\% C.L.) \cite{Garcia-Gallego:2025kiw}.

\begin{figure} 
    \centering
    \includegraphics[width=0.8\linewidth]{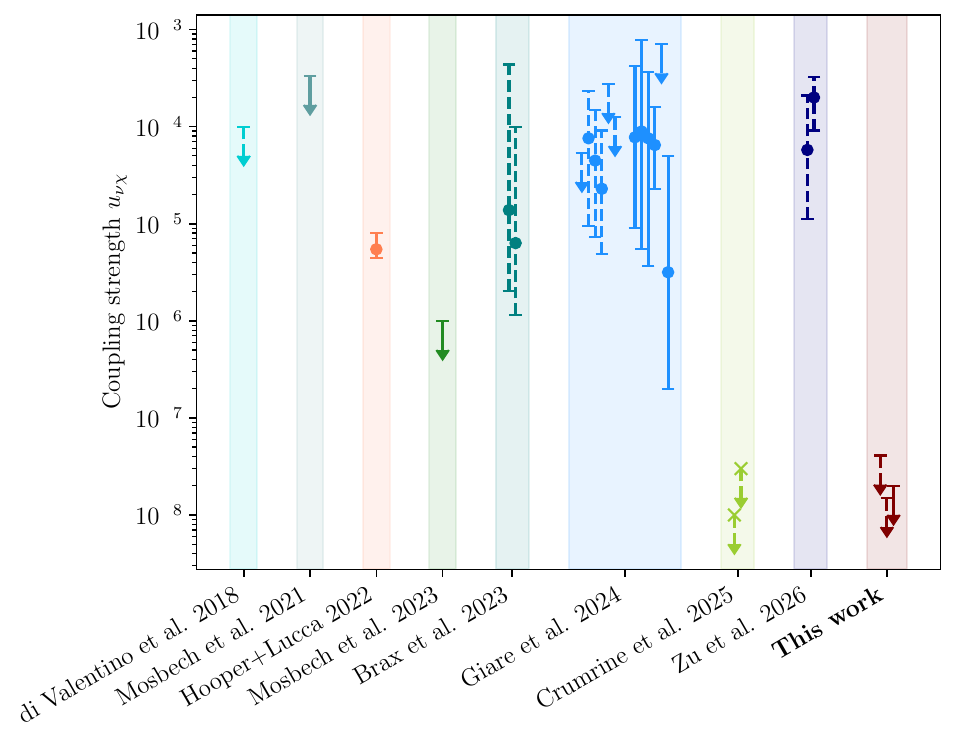}
    \caption{Overview of selected limits and preferred values from the literature (di Valentino et al. 2018~\cite{DiValentino:2017oaw}, Mosbech et al. 2021~\cite{Mosbech:2020ahp}, Hooper+Lucca 2022~\cite{Hooper:2021rjc}, Mosbech et al. 2023~\cite{Mosbech:2022nkk}, Brax et al. 2023~\cite{Brax:2023rrf}, Giare et al. 2024~\cite{Giare:2023qqn}, Crumrine et al. 2025~\cite{Crumrine:2024sdn}, Zu et al. 2026~\cite{Zu:2025lrk}), as well as those presented in this work. Dashed lines refer to limits for massless neutrinos, while solid lines are for neutrinos of non-zero mass. Ref.~\cite{Crumrine:2024sdn} is the only one obtaining constraints comparable to this work, which were inferred from Milky Way satellites. However, these limits were obtained from a linear mapping between WDM and iDM models rather than a full simulation of the iDM model, which adds additional uncertainty to the reported limits.}
    \label{fig:all_limits}
\end{figure}

\section{Conclusion}
\label{sec:conclusion}
We have presented a set of direct cosmological hydrodynamical simulations of models with DM-neutrino interactions, spanning a range of thermal histories and interaction strengths. We have extracted the 1D Lyman-$\alpha$ flux power spectrum from these simulations, and used the results to train an emulator following the same methodology as in Ref.~\cite{Garcia-Gallego:2025kiw}. We used the new emulator to perform a few MCMC analyses with high resolution Lyman-$\alpha$ forest data from HIRES/Keck and UVES/VLT~\cite{Boera:2019}. We inferred an upper bound on the DM-neutrino interaction strength, $\udm \leq 1.5\times 10^{-8}$ (95\% C.L.), based on a prior on the optical depth given by CMB observations. Our constraints degrade only slightly when using alternative priors. The simulations were performed using the approximation of massless neutrinos, but we find that the limit can be directly mapped to $\udm \leq 2.0\times 10^{-8}$ in the case of three neutrinos degenerate in mass with $\sum m_{\nu}=0.06$ eV. This mapping relies on a nearly perfect correspondence at the level of linear power spectra.

Our results provide the strongest bound to date on the interaction strength $\udm$ inferred directly from dedicated simulations. They confidently exclude the hint for a non-zero interaction strength inferred from earlier Lyman-$\alpha$~\cite{Hooper:2021rjc}, CMB~\cite{Brax:2023rrf,Giare:2023qqn}, and large scale structure~\cite{Zu:2025lrk} data. We show a graphical comparison of our results with other limits and detection hints reported in the literature in Fig.~\ref{fig:all_limits}. Our bounds are significantly stronger than all previous ones excepted those from Ref.~\cite{Crumrine:2024sdn} and \cite{Cline:2023tkp}, which are of comparable magnitude. However, the former were inferred from a linear mapping to simulated WDM models rather than full simulations of iDM models, and the latter relies on a model for the dark matter halo and active nucleus of NGC 1068. Therefore, they involve additional layers of uncertainty.

Our suite of full simulations of iDM models allowed us to assess the accuracy of approximate methods used in the literature to convert WDM limits into limits on other DM models with the same level of small-scale structure suppression, such as iDM. We found that these methods, including the $\delta A$ and half-mode criteria, slightly overestimate the constraining power of Lyman-$\alpha$ data for iDM models.

\acknowledgments
The authors gratefully acknowledge the computing time provided to them at the NHR Center NHR4CES at RWTH Aachen University (project number p0021792). This work is funded by the Federal Ministry of Education and Research, and the state governments participating on the basis of the resolutions of the GWK for national high performance computing at universities (\url{www.nhr-verein.de/unsere-partner}).
MV is supported by INFN INDARK, IDEAS SISSA and the INAF theory "Cosmological Investigation of the Cosmic Web" grants. JL and MM acknowledge support from the DFG through the grants LE 3742/9-1 and  KA 4662/4-1.

\bibliography{bibliography}

\appendix
\section{Triangle Plot}
\label{sec:triangle}
We present here the triangle plot showing 2D-posteriors for the interaction strength $\udm$ as well as the astrophysical parameters.

\begin{figure}[h] 
    \centering
    \includegraphics[width=\linewidth]{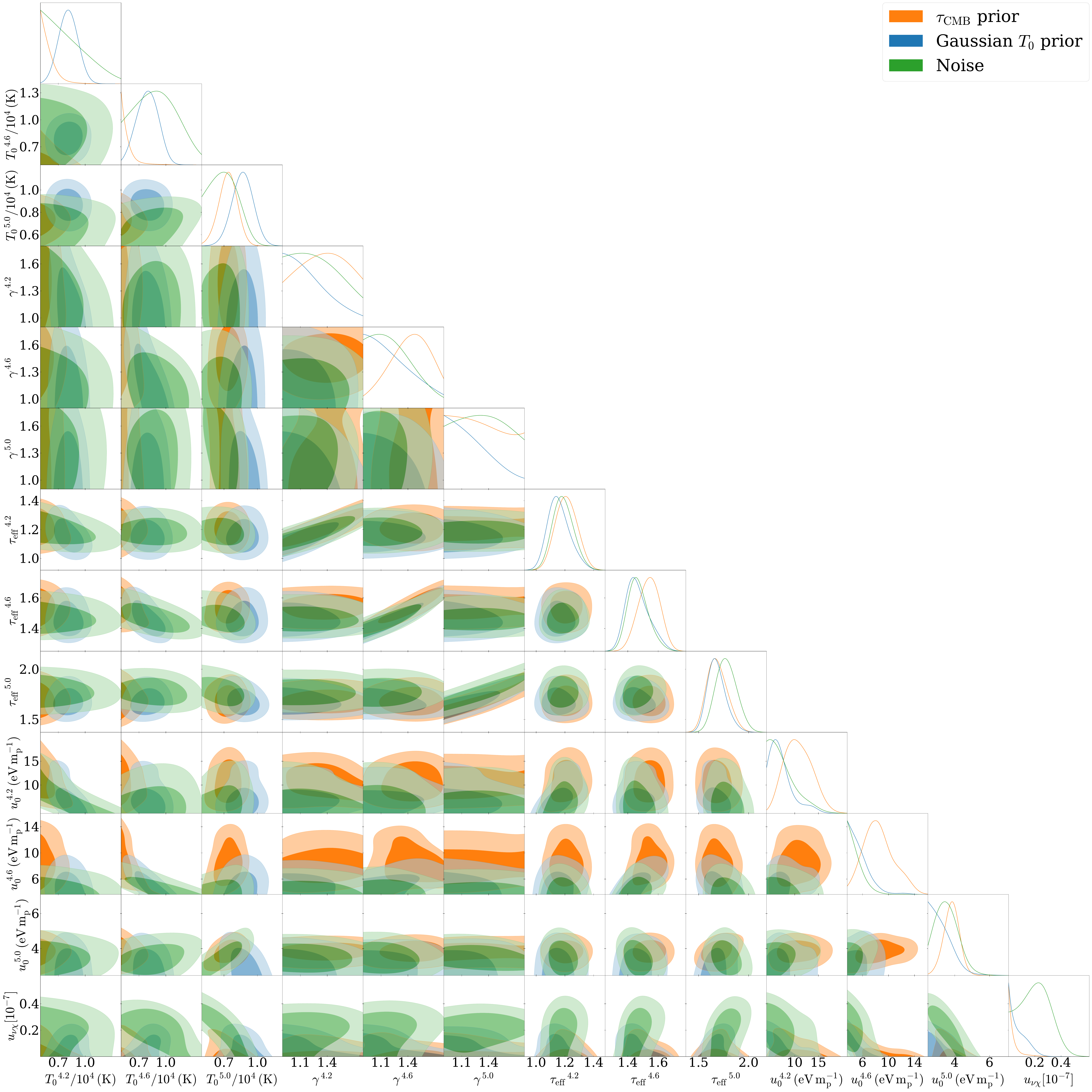}
    \caption{Triangle plot for the baseline analysis ($\tau_{\rm{CMB}}$ priors), Gaussian $T_0$ priors, and Noise analysis (with $\tau_{\rm{CMB}}$ priors) in orange, blue and green, respectively.}
    \label{fig:corner}
\end{figure}

\end{document}